\documentclass[authoryear,preprint,review,12pt]{elsarticle}
\usepackage{amssymb}
\usepackage{epsfig}
\usepackage{amssymb}
\usepackage{setspace}
\usepackage{graphics}
\usepackage{epsfig}
\usepackage{color}
\usepackage{graphicx}
\usepackage{amsmath,bbm,amssymb}
\usepackage{amsthm}
\usepackage{comment}
\usepackage{rotating}
\usepackage{lscape}

\newtheorem{theorem}{Theorem}[section]
\newtheorem{lemma}{Lemma}[section]
\newtheorem{corollary}{Corollary}[section]

\journal{Journal of Statistical Planning and Inference}

\begin{document}

\begin{frontmatter}
\title{\textbf{Robustness of Optimal Designs for $2^2$ Experiments with Binary Response}}
\author{Jie Yang\fnref{label1}}
\author{Abhyuday Mandal\corref{cor1}\fnref{label2,label3}}
\author{Dibyen Majumdar\fnref{label1}}
\address[label1]{Department of Mathematics, Statistics, and Computer Science, University of Illinois at Chicago, Chicago, IL 60607-7045, USA}
\address[label2]{Department of Statistics, University of Georgia, Athens, GA 30602-7952, USA}
\fntext[label3]{This research is partially supported by NSF research grant DMS-0905731.}
\cortext[cor1]{Corresponding author at: Department of Statistics, University of Georgia, Athens, GA, 30602 - 7952, USA. E-mail: amandal@stat.uga.edu.}

\begin{abstract}
We consider an experiment with two qualitative factors at $2$ levels each and a binary response, that follows a generalized linear model. In Mandal, Yang and Majumdar (2010) we obtained basic results and characterizations of locally $D$-optimal designs for special cases. As locally optimal designs depend on the assumed parameter values, a critical issue is the sensitivity of the design to misspecification of these values. In this paper we study the sensitivity theoretically and by simulation, and show that the optimal designs are quite robust. We use the method of \textit{cylindrical algebraic decomposition} to obtain locally $D$-optimal designs in the general case.
\end{abstract}


\end{frontmatter}

\singlespacing

\section{Introduction}

\bigskip\noindent We consider experiments with two qualitative factors at two levels each with binary response, and investigate the robustness of the $D$-optimal allocation of replicates. Binary responses are usually modelled using generalized linear models (GLMs). GLMs have been used widely for modelling the mean response both for discrete and continuous random variables with an emphasis on categorical response. Although the methods of analyzing data using these models have been discussed in depth in the literature (McCullagh and Nelder (1989), Agresti (2003)), only a limited number of results are available for the choice of optimal design of experiments under GLMs (Khuri et al. (2006)). Recently Mandal, Yang and Majumdar (2010) obtained some theoretical results for locally D-optimal $2^2$ designs in some special cases. Recall that Fisher's Information matrix contains the unknown parameters, and so does the D-optimality criterion. As we consider the locally optimal designs, a critical question is how to choose the initial values of the unknown parameters, and also to investigate the robustness of the optimal designs with respect to misspecification of the values of the parameters, an issue which was briefly mentioned but not explored in details in our earlier parer. In this paper, we obtain theoretical as well as numerical results on the optimality of $2^2$ designs. We investigate thoroughly the robustness of designs, with particular emphasis on the uniform ($i.e.$, equireplicate) design.

\bigskip\noindent The optimality criterion can be written in terms of the variances, or information, at each of the $2^{2}$ points. Note that these variances depend on the parameters through the link function. It turns out that the $D$-optimal design can be quite different from the uniform design, especially when at least two of these variances are far from each other.

\bigskip\noindent Our results can be described as follows. If the experimenter has reliable knowledge of the variances then the design obtained by using the cylindrical algebraic decomposition (CAD) technique discussed later results in a highly efficient design. If the experimenter has some approximate knowledge of those variances, then using our results, efficient designs can be obtained. In the absence of any prior idea of the variances our recommendation is to use the uniform design, which is the most robust one in general. It may be noted that for applications where a $D$-optimal design cannot be used, it can still serve as a benchmark to evaluate other designs.

\bigskip\noindent For illustration suppose we have a linear predictor involving two factors, namely, family history ($x_1$) and physical exercise ($x_2$), which leads to the linear predictor $\eta=\beta_0+\beta_1x_1+\beta_2x_2$. If we have sound reasons to believe that family history is a much more dominant factor than exercise, and on the basis of prior knowledge, an initial choice of parameters $(\beta_0,\beta_1,\beta_2) = (2,2,0.05)$ is reasonable, then the optimal allocation of 100 experimental units using the CAD method will be $(6,28,33,33)$. One the other hand, if the practitioner decides to use the popular uniform design, then the relative loss of efficiency is approximately 5\%.

\bigskip\noindent This paper is organized as follows. In section 2 we give the preliminary setup and some relevant results from Mandal, Yang and Majumdar (2010). The cylindrical algebraic decomposition technique to solve the general $D$-optimality problem for $2^{2}$ experiment will be discussed in section 3. In section 4 we discuss robustness of the $D$-optimal designs both theoretically and numerically. We study the robustness of uniform design as well as a real experiment in section 5 and conclude with some remarks in section 6. All proofs are relegated to the appendix.

\section{$D$-optimal $2^2$ Designs: Preliminary Setup}

\bigskip\noindent Consider a $2^{k}$ experiment, i.e., an experiment with $k$ explanatory variables at $2$ levels each. Suppose the proportion of units allocated to the $i$th experimental condition is $p_i$, such that $p_{i}\geqslant 0,$ $i=1,\ldots,2^{k}$, and $\sum p_i = 1$. In this paper we consider the problem of finding the ``optimal'' $p_i$'s. Let $\eta $ be the linear predictor for the chosen model. For instance, in a $ 2^{3} $ experiment, $\eta =\beta _{0}+\beta _{1}x_{1}+\beta _{2}x_{2}+\beta _{3}x_{3}+\beta _{23}x_{2}x_{3}$ represents a model that includes all the main effects and the two-factor interaction of factors $2$ and $3$. The aim of the experiment is to obtain inferences about the parameter vector of factor effects $\beta ;$ in the preceding example, $\beta =\left( \beta _{0},\beta _{1},\beta _{2},\beta _{3},\beta _{23}\right) ^{\prime }.$

\bigskip\noindent In this paper, we consider the $2^{2}$ experiment with main-effects model, which gives $ \eta =\beta _{0}+\beta _{1}x_{1}+\beta _{2}x_{2}$ and $\beta =\left( \beta _{0},\beta _{1},\beta _{2}\right) ^{\prime }$. Here the mean response $\mu$ is connected to the linear predictor $\eta$ by the link function (McCullagh and Nelder (1989)). 

\bigskip\noindent The maximum likelihood estimator of $\beta $ has an asymptotic covariance matrix that is the inverse of $ nX^{\prime }WX$, where $W=diag\left( w_{1}p_{1},...,w_{2^{k}}p_{2^{k}}\right)$, $w_{i} = \left( \frac{d\mu _{i}}{ d\eta _{i}}\right) ^{2}/\left( \mu _{i}(1-\mu _{i})\right) $ and $X$ is the ``model matrix''. For example, for a $2^{2}$ experiment with main-effects model, $X = ((1, 1, 1, 1)'$, $(1, 1, -1, -1)'$, $(1, -1,$ $1,$ $-1)')$. The {\it $D$-optimality} criterion maximizes $\left\vert X^{\prime }WX\right\vert^{1/p}$ where $p$ is the number of parameters in the model; in this case $p=3$. In the rest of this section, we will give some basic ideas and results developed in Mandal, Yang and Majumdar (2010).

\bigskip\noindent For the $2^2$ experiment with main-effects model, the optimization problem maximizing $\left\vert X^{\prime }WX\right\vert^{1/p}$ reduces to maximizing $$det(\mathbf{w},\mathbf{p}) = 16w_1w_2w_3w_4L(\mathbf{p}),$$
where ${\mathbf w}=(w_1,w_2,w_3,w_4)'$, ${\mathbf p}=(p_1,p_2,p_3,p_4)'$,
\begin{equation}\label{simplifiedprob}
L({\mathbf p})=v_4p_1p_2p_3+v_3p_1p_2p_4+v_2p_1p_3p_4+v_1p_2p_3p_4,
\end{equation}
and $v_i = 1/w_i,\ i=1,2,3,4$. Assuming $v_i>0$ for $i=1,2,3,4$, we consider the problem of maximizing $L(\mathbf{p})$ over all vectors $\mathbf{p}$ with $p_i \geq 0$ and $\sum_ip_i=1$. Although the objective function (\ref{simplifiedprob}) is elegant, an analytic solution with general $v_i>0$ is not available. Mandal, Yang and Majumdar (2010) proved the following results.
\begin{lemma}\label{lemma:1} If $v_1 > v_2$, then any solution maximizing $L(\mathbf{p})$ must satisfy $p_1\leq p_2$;
if $v_1=v_2$, then any solution must satisfy $p_1=p_2$.
\end{lemma}

\begin{theorem}\label{thm:1} $L(\mathbf{p})$ has a unique maximum at ${\mathbf p} = (0, 1/3, 1/3, 1/3)$ if and only if $v_1\geq v_2+v_3+v_4$.
\end{theorem}

\noindent Note that this does not correspond to a complete $2^2$ experiment, rather it corresponds to a design supported only on three points, which is saturated for the main effects plan $\eta =\beta_0 + \beta_1x_{1} + \beta_2 x_{2}$. Theorem~\ref{thm:1} reveals that the $D$-optimal design is saturated if and only if
$2\max_i v_i \geq v_1 + v_2 + v_3 + v_4$, that is,
\begin{equation}\label{boundarycondition}
\frac{2}{\min\{w_1,w_2,w_3,w_4\}} \geq \frac{1}{w_1} + \frac{1}{w_2} + \frac{1}{w_3}
+\frac{1}{w_4}.
\end{equation}
We call (\ref{boundarycondition}) the {\it saturation condition}. This condition will play a crucial role in our robustness study. We defer the discussion on the importance of condition until section~4.
We also need the two corollaries listed below in the robustness study.

\begin{corollary}\label{corollary:1}
Suppose $v_2=v_3=v_4=v$ and $v_1 < 3v$.  Then the solution maximizing (\ref{simplifiedprob}) is
\[
    p_1=\frac{3v - v_1}{9v-v_1},\> p_2=p_3=p_4=\frac{2v}{9v-v_1}
\]
with the maximum $L=4v^3/(9v-v_1)^2$.
\end{corollary}

\begin{corollary}\label{corollary:2}
Suppose $v_1=v_2=u$, $v_3=v_4=v$, and $u>v$. Then the solution maximizing (\ref{simplifiedprob}) is
\[
    p_1=p_2=\frac{2u-v-d}{6(u-v)},\> p_3=p_4=\frac{u-2v+d}{6(u-v)}
\]
with the maximum $L=\left.(2u-v-d)(u-2v+d)(u+v+d)\right/\left[108(u-v)^2\right]$, where $d = \sqrt{u^2 - uv + v^2}$.
\end{corollary}

\section{Exact Solutions Using Cylindrical Algebraic Decomposition}

\bigskip \noindent Since analytic solutions for the optimization problem in (\ref{simplifiedprob}) is not available, in this section we will investigate computer-aided exact solution. One option is to use the Lagrange multipliers or the Karush-Kuhn-Tucker (KKT) conditions (Karush (1939), Kuhn and Tucker (1951)). It leads to intractable polynomial equations. Another option is to use numerical search algorithms such as Nelder-Mead, quasi-Newton, conjugate-gradient, or simply a grid search (for a comprehensive reference, see Nocedal and Wright (1999)). Those numerical methods are computational intensive in general when an accurate solution is needed.
We suggest using the {\it cylindrical algebraic decomposition} (CAD) algorithm to find the exact global solution.

\bigskip\noindent Fotiou et al. (2005) gave detailed description of using CAD for general constrained optimization problems. Our optimization problem (\ref{simplifiedprob}) is associated with the so-called {\it boolean expression}:
$$(L_3-f\geq 0)\bigwedge
(p_1\geq 0)\bigwedge
(p_2\geq 0)\bigwedge
(p_3\geq 0)\bigwedge
(p_1+p_2+p_3\leq 1)$$
where $L_3=L(p_1,p_2,p_3,1-p_1-p_2-p_3)$,
and $f$ is a new parameter indicating the value of the objective function. Given specific values of $v_1, v_2, v_3, v_4$,
the CAD can represent the feasible domain of $(f,p_1,p_2,p_3)$ in ${\mathbb R}^4$ as a finite union of disjoint cells.
Each cell takes the form
\[
\left\{
\begin{array}{c}
(f,p_1,p_2,p_3)\\
\in {\mathbb R}^4
\end{array}
 \left|
 \begin{array}{l}
 f=a_0\mbox{ or } a_0 < f < b_0,\\
 p_1 = g_1(f)\mbox{ or } g_1(f) < p_1 < h_1(f),\\
 p_2 = g_2(f,p_1)\mbox{ or } g_2(f,p_1) < p_2 < h_2(f,p_1),\\
 p_3 = g_3(f,p_1,p_2)\mbox{ or }g_3(f,p_1,p_2) < p_3 < h_3(f,p_1,p_2)
 \end{array}
 \right.
\right\}
\]
for some constants $a_0,b_0$ and some functions $g_i,h_i$, $i=1,2,3$. Since $f$ indicates the value of the objective function $L$ (or $L_3$), the cell with greatest $f$ provides us the maximum of $L$. We will illustrate the method with an example.

\bigskip \noindent Suppose $v_1=1, v_2=2, v_3=3, v_4=4$. Using the software {\tt Mathematica}, we obtain that the maximum of $L$ based on CAD is the negative first root of equation $-96 + 800 x + 5220 x^2 - 19035 x^3 + 2187 x^4=0$ and
\begin{itemize}
\item[] $p_1$ is the $4$th root of equation $-2 - 13 x + 18 x^2 + 126 x^3 + 54 x^4=0,$\vspace{-.1in}
\item[] $p_2$ is the $2$nd root of equation $-2 + 2 x + 28 x^2 - 39 x^3 + 9 x^4=0,$\vspace{-.1in}
\item[] $p_3$ is the $2$nd root of equation $-3 + 13 x + 2 x^2 - 26 x^3 + 6 x^4=0$.
\end{itemize}
Here the numerical solution is $\max L=0.1645$ with $p_1=0.3112, p_2=0.2849, p_3=0.2508, p_4=0.1531$.

\bigskip \noindent Note that the CAD algorithm can be used to find exact solution for general $v_1, v_2, v_3, v_4$, although an explicit formula is not available. This technique will be used in the next section for robustness study.

\section{Robustness for $2^2$ Designs}

\bigskip \noindent Since locally optimal designs depend on the assumed values of the parameters, it is important to study the robustness of the designs to these values.  For experiments where there is no basis for making an informed choice of the assumed values, the natural design choice is the uniform design.  In this section, we study the robustness of the optimal design for misspecification of assumed values.

\subsection{Robustness for misspecification of $w$}

\bigskip \noindent Suppose ${\mathbf w_t} = (w_{t1}, w_{t2}, w_{t3}, w_{t4})$ is the true ${\mathbf w}$, and ${\mathbf w_c} = (w_{c1}, w_{c2}, w_{c3}, w_{c4})$ is the chosen (assumed) ${\mathbf w}$.
Let ${\mathbf p_t} = (p_{t1}, p_{t2}, p_{t3}, p_{t4})$ and ${\mathbf p_c} = (p_{c1}, p_{c2}, p_{c3}, p_{c4})$ be the optimal designs corresponding  to ${\mathbf w_t}$ and ${\mathbf w_c}$, respectively.
The relative loss of efficiency of choosing ${\mathbf w_c}$ instead of ${\mathbf w_t}$ may be defined as
\begin{eqnarray}\label{eqn:R}
R(t,c)=\frac{det(\mathbf{w}_t,\mathbf{p}_{t})^{1/3}-det(\mathbf{w}_{t},\mathbf{p}_c)^{1/3}}{det(\mathbf{w}_{t},\mathbf{p}_{t})^{1/3}},
\end{eqnarray}
where the notation $det(\mathbf{w},\mathbf{p})$ was defined prior to expression  (\ref{simplifiedprob}). Note that $R(t,c)$ in equation~(\ref{eqn:R}) remains invariant under scalar multiplication of determinants. Now let us define the maximum relative loss of efficiency as
\begin{eqnarray}\label{eqn:max}
R_{max}(c) = \max_{t}\Big\{R(t,c)\Big\}.
\end{eqnarray}
This maximum will correspond to the worst case scenario. This tells us, for each $\mathbf{w}$, how bad the design can perform if we do not choose the $\mathbf{w}$ correctly.

\begin{figure}\caption{plot of $w$ versus $\pi$}\label{fig1}
\begin{center}
 \includegraphics[height=2in,width=2.2in,angle=0]{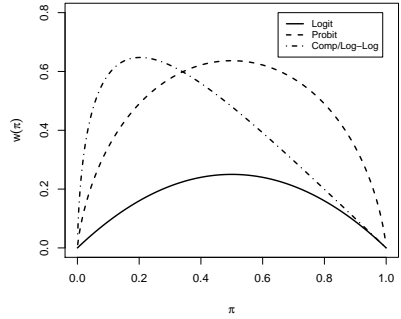}
\end{center}
\end{figure}

\bigskip \noindent For a binary response, we consider the commonly used link functions including logit, probit, log-log and complimentary log-log links. Figure~\ref{fig1} illustrates the range of $w$ for specific link functions. The logit link corresponds to $0\le w \le 0.25$ whereas for the probit link $0\le w \le 2/\pi$ and for the (complementary) log-log links $0\le w \le 0.648$. It should also be noted that the $w$-curve is symmetric for logit and probit links but asymmetric for the (complementary) log-log link. To examine the robustness for mis-specification of $w$ for different links, we assume $0 < \alpha \leq w \leq \beta$ for some constants $\alpha$ and $\beta$, since $w=0$ leads to trivial cases.

\bigskip \noindent Fixing a chosen ${\mathbf w_c} = (w_{c1}, w_{c2}, w_{c3}, w_{c4})$, let $v_{ci}=1/w_{ci}$, $i=1,2,3,4$. Without any loss of generality, we assume $v_{c1} \leq v_{c2} \leq v_{c3} \leq v_{c4}$. It follows from Lemma~\ref{lemma:1} that $p_{c1} \geq p_{c2} \geq p_{c3} \geq p_{c4}$. For the true ${\mathbf w}_t$, define ${\mathbf v}_t = (v_{t1}, v_{t2}, v_{t3}, v_{t4})$ with $v_{ti}=1/w_{ti}$, $i=1,2,3,4$. In practice, the experimenter might have some rough idea about the range of the parameter values and our next Theorem specifies the worst possible performance of a chosen design for that assumed range.

\begin{theorem}\label{thm:2}
Suppose $0<a\leq v_{c1} \leq v_{c2} \leq v_{c3} \leq v_{c4} \leq b$
and $a\leq v_{ti} \leq b$, $i=1,2,3,4$.
\begin{itemize}
\item[{\bf (i)}]
If $v_{c4} \geq v_{c1} + v_{c2} + v_{c3}$, then
$R_{\max}(c)
=1-\frac{1}{3\theta}\left(\frac{9\theta-1}{2}\right)^{2/3}$,
where $\theta=b/a \geq 3$, and the maximum can only be attained at
${\mathbf v}_t=(b,b,b,a)$.
\item[{\bf (ii)}]
If $v_{c4} < v_{c1} + v_{c2} + v_{c3}$, then
$R_{\max}(c)$ can only be attained at ${\mathbf v}_t=(b, a, a, a)$,
$(b,b,a,a)$, or $(b,b,b,a)$.
\end{itemize}
\end{theorem}


\bigskip \noindent If $w$ can be as small as $0$, for example, for logistic regression,
$0\leq w \leq 0.25$, then the upper bound $b$ for
$v$ will go to $\infty$.
As a direct conclusion from Theorem~\ref{thm:2}, we obtain

\begin{corollary}\label{corollary:3}
Suppose $0<a\leq v_{c1} \leq v_{c2} \leq v_{c3} \leq v_{c4} < \infty$
and $a\leq v_{ti} < \infty$, $i=1,2,3,4$.
\begin{itemize}
\item[{\bf (i)}]
If $v_{c4} \geq v_{c1} + v_{c2} + v_{c3}$, then
$R_{\max}(c) = 1$.
\item[{\bf (ii)}]
If $v_{c4} < v_{c1} + v_{c2} + v_{c3}$, then
$R_{\max}(c) = 1 - 3\left(p_{c2}p_{c3}p_{c4}\right)^{1/3}$.
\end{itemize}
\end{corollary}

\bigskip \noindent To find out the most robust design in terms of
maximum relative loss, we need explicit formulas
of $R_{\max}(c)$ for case (ii) of Theorem~\ref{thm:2}. To simplify the notations, let
$q_4  = p_{c1} p_{c2} p_{c3}$, $q_3  = p_{c1} p_{c2} p_{c4}$, $q_2  = p_{c1} p_{c3} p_{c4}$, $q_1  = p_{c2} p_{c3} p_{c4}$.
Then $q_4  \geq q_3  \geq q_2  \geq q_1 $. Let
$$
Q_c(v_{t1}, v_{t2}, v_{t3}, v_{t4})
=
\frac{v_{t4} q_4  + v_{t3} q_3  + v_{t2} q_2  + v_{t1} q_1 }
{v_{t4} p_{t1} p_{t2} p_{t3} + v_{t3} p_{t1} p_{t2} p_{t4} + v_{t2} p_{t1} p_{t3} p_{t4} + v_{t1} p_{t2} p_{t3} p_{t4}}~.
$$
Then $R(t,c) = 1-Q_c(v_{t1}, v_{t2}, v_{t3}, v_{t4})^{1/3}$
and the $R_{\max}(c)$ in case (ii) of Theorem~\ref{thm:2} is
$$\max\left\{1-Q_c(b, a, a, a)^{1/3}, \> 1-Q_c(b,b,a,a)^{1/3},\> 1-Q_c(b,b,b,a)^{1/3}\right\}.
$$
Let $\theta = b/a \geq 1$ and $\rho=\sqrt{\theta^2-\theta+1}$.  Based on Corollary~\ref{corollary:1}
and Corollary~\ref{corollary:2},
\begin{eqnarray*}
Q_c(b,a,a,a) &=& Q_c(\theta,1,1,1) =
\left\{
\begin{array}{ll}
\frac{27}{\theta}(\theta q_1 + q_2 + q_3 + q_4), & \mbox{ if }\theta\geq 3\\
&\\
\frac{(9-\theta)^2}{4}(\theta q_1  + q_2 + q_3 + q_4), & \mbox{ if }1\leq \theta < 3
\end{array}
\right.
\\
Q_c(b,b,a,a) &=& Q_c(\theta,\theta,1,1) = \frac{108(\theta-1)^2(\theta q_1 + \theta q_2 + q_3 + q_4)}{(2\theta-1-\rho)(\theta-2+\rho)(\theta+1+\rho)}\\
Q_c(b,b,b,a) &=& Q_c(\theta,\theta,\theta,1) = \frac{(9\theta-1)^2}{4\theta^3}(\theta q_1  + \theta q_2  + \theta q_3 + q_4 ).
\end{eqnarray*}

\bigskip\noindent
Note that $R_{\max}(c)$ is actually a function of ${\mathbf p}_c=(p_{c1}, p_{c2},
p_{c3}, p_{c4})$.

\begin{theorem}\label{thm:3}
Suppose $v_{ci}, v_{ti} \in [a,b]$, $i=1,2,3,4$, $0 < a \leq b$.
Then $R_{\max}(c)$ attains its minimum if and only if
${\mathbf p}_c = (1/4, 1/4, 1/4, 1/4)$, which is the uniform design.
\end{theorem}

\bigskip\noindent
In other words, the uniform design is the most robust one in terms of
the maximum of relative loss of efficiency.
Note that the conclusion of Theorem~\ref{thm:3} is still true even if $[a,b]$
is replaced with $[a,\infty)$.

\subsection{Simulation study}

\bigskip \noindent To study the robustness measured by percentiles of $\{R(t,c)\}$ other than the maximum $R_{\max}(c)$, we randomly select 1000 vectors $\mathbf{w}_i=(w_{i1}, w_{i2}, w_{i3}, w_{i4})$, $i=1,2,\ldots, 1000$. For the logit link, $0\le w_i \le 0.25$. If we randomly select $w_i$'s between $0$ and $0.25$, the chance of getting a saturated design can be as high as 48\% when some $w_i$ is close to $0$ and the condition of Theorem~\ref{thm:1} applies. We try to skip the cases that give a saturated design since in those cases both the exact solution and robustness are clearly known. So here we consider $w\ge 0.05$ only. Then the chance of saturated design drops to 6\% for uniformly distributed $w_i$'s. So, for the logit link, we consider $0.05\le w \le 0.25$ and for the other links, $0.05\le w \le 0.65$.

\bigskip \noindent Suppose ${\mathbf w_t} = (w_{t1}, w_{t2}, w_{t3}, w_{t4})$ is the true ${\mathbf w}$, and ${\mathbf w_c} = (w_{c1}, w_{c2}, w_{c3}, w_{c4})$ is the chosen (assumed) ${\mathbf w}$. We consider 1000 cases. In our robustness study, each one of the 1000 ${\mathbf w}$'s is chosen in turn as ${\mathbf w_c}$ and the remaining 999 cases are regarded as ${\mathbf w_t}$ respectively. We use CAD to determine the optimal designs ${\mathbf p_t} = (p_{t1}, p_{t2}, p_{t3}, p_{t4})$ and ${\mathbf p_c} = (p_{c1}, p_{c2}, p_{c3}, p_{c4})$ corresponding  to ${\mathbf w_t}$ and ${\mathbf w_c}$, respectively.

\begin{figure}\caption{Robustness study : performance of the ``worst 1\%'' design}\label{fig5}
\begin{center}
\begin{tabular}{cc}
  \includegraphics[height=2.2in,width=2.2in,angle=0]{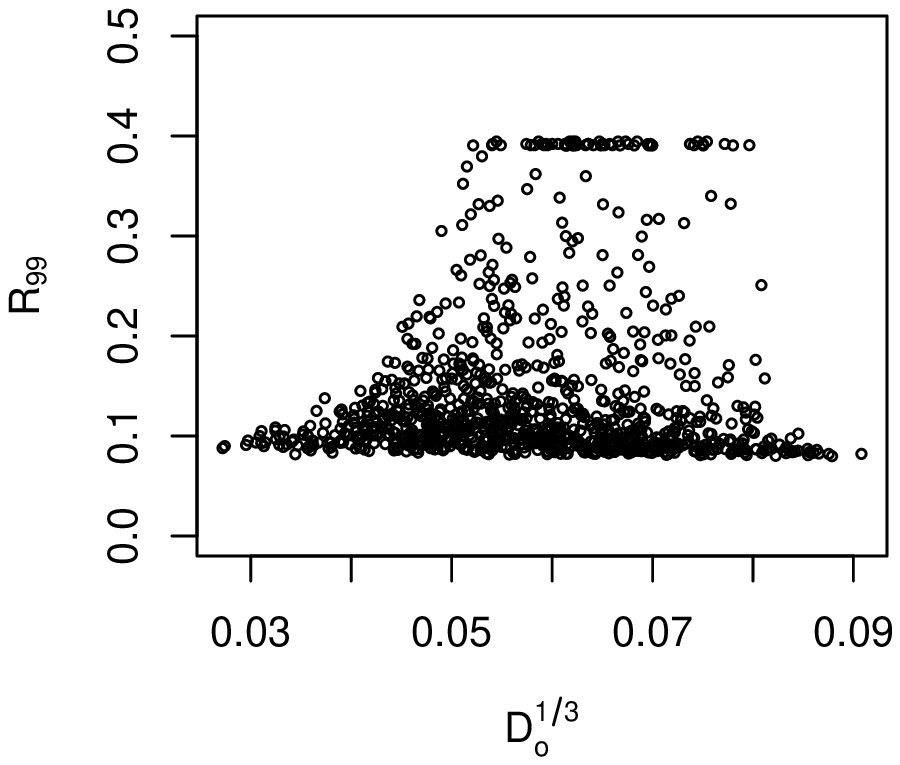} &  \includegraphics[height=2.2in,width=2.2in,angle=0]{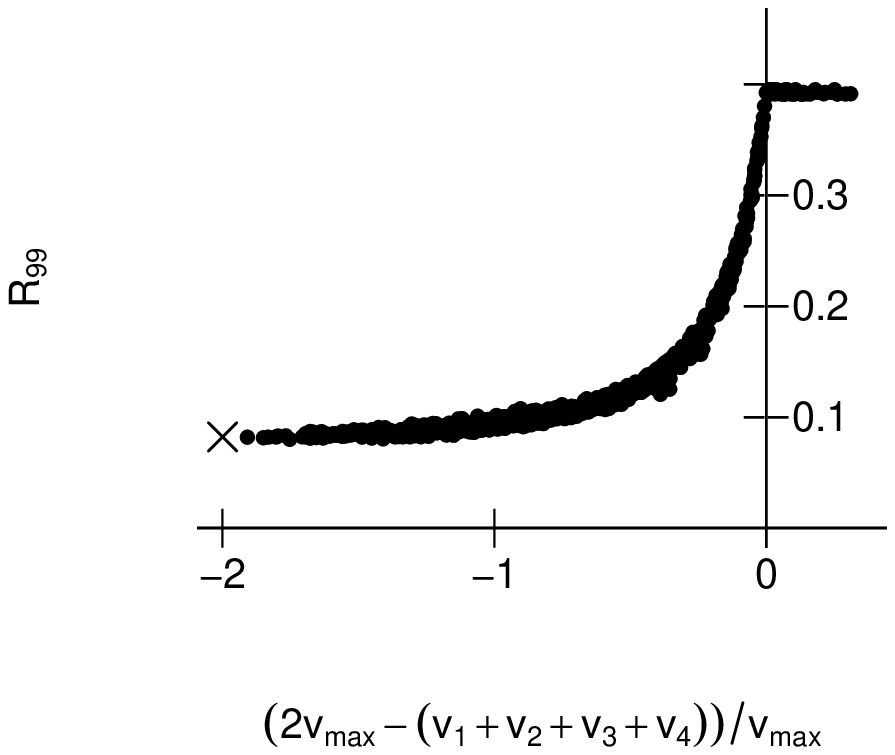}\\[-11pt]
\end{tabular}
\end{center}
\end{figure}

\bigskip \noindent For the numerical computations in this section, we consider the upper 99$^{\rm th}$ percentile of $\{R(t,c)\}_{t=1}^{1000}$ and denote it by $R_{99}(c)$. This corresponds to the worst $1\%$  case scenario. The left panel of Figure~\ref{fig5} illustrates that this relative loss will range roughly between 0.1 and 0.4, whereas the right panel helps us identify those ${\mathbf w}$'s with non-robust optimal solution ${\mathbf p}$. The horizontal axis corresponds to the distance between $v_{\rm max}$ and $\sum v_i - v_{\rm max}$ divided by $v_{\rm max}$, where $v_{\rm max} =\max\{v_1, v_2, v_3, v_4\}$. The vertical axis is our robustness measurement $R_{99}$. There is a clearly positive association between the relative loss and the distance.  We have examined the other quantiles such as the 25$^{\rm th}$ quantile, median, 75$^{\rm th}$ quantile, and 95$^{\rm th}$ quantile of $R(t,c)$. The patterns are similar for all of them. From this, we conclude that the locally $D$-optimal designs are quite robust and the farther the ${\mathbf w}$'s are from the saturation condition (\ref{boundarycondition}), the smaller is the relative loss of efficiency. It is interesting to note that the left most point (denoted by $\times$) on the right panel of Figure~\ref{fig5} corresponds to the uniform design. It can be verified that the standardized distance $(2v_{\rm max}-\sum v_i)/v_{\rm max}$ attains its minimum $-2$ if and only if $v_1 = v_2 = v_3 = v_4$ which leads to the uniform design. While the left panel of Figure~\ref{fig5} indicates that the performance of the ``worst 1\%'' designs is not too bad in terms of robustness, the right panel (as well as figures of other quantiles, not shown here) clearly demonstrates the significance of the saturation condition. The points with $R_{99}$ values greater than 0.15 either satisfy or almost satisfy the saturation condition. This figure also suggests that the uniform design is highly robust.

\bigskip \noindent Similar results have been obtained for probit and (complementary) log-log links.

\section{Robustness of Uniform Design}

\bigskip \noindent If the experimenter is unable to make an informed choice of the assumed values for local optimality, the natural design choice is the uniform design ${\mathbf p}_{u}=(1/4,1/4,1/4,1/4).$ The relative loss of efficiency of ${\mathbf p}_{u}$ with respect to the true ${\mathbf w} = (w_1, w_2, w_3, w_4)$ is:
\begin{equation*}
R_{u}({\mathbf w})=\frac{\det \left( {\mathbf w}, {\mathbf p}_t\right) ^{1/3}-\det \left( {\mathbf w},{\mathbf p}_{u}\right) ^{1/3}}{\det \left( {\mathbf w}, {\mathbf p}_{t}\right) ^{1/3}}
\end{equation*}
It can be shown that
\begin{equation*}
R_{u}({\mathbf w})= 1 - \frac{1}{4}\left(\frac{ v_1+v_2+v_3+v_4}{L({\mathbf p}_t)}\right)^{1/3},
\end{equation*}%
where $v_i = 1/w_i$, ${\mathbf p}_t$ is the optimal design under ${\mathbf w}$, and $L({\mathbf p}_t)$ is defined in (\ref{simplifiedprob}).


\subsection{Maximum relative loss of uniform design}

\bigskip\noindent We denote by $R_{\rm max}^{(u)}=\underset{{\mathbf w}}{\max }$ $R_{u}({\mathbf w})$ the maximum loss of efficiency of the uniform design. The following theorem formulates $R_{\rm max}^{(u)}$ with different values of ${\mathbf w}$'s and generalizes Theorem~4.1.5 of Mandal, Yang and Majumdar (2010).

\begin{theorem}\label{thm:42}
Suppose $0<\alpha\leq w_i\leq \beta$, $i=1,2,3,4$. Let $\theta=\beta/\alpha \geq 1$. Then
\[
R_{\rm max}^{(u)}=\left\{
\begin{array}{ll}
1 - \frac{3}{4}\left(1+\frac{3}{\theta} \right)^{1/3}, & \mbox{ if }\theta\geq 3\\
1 - \frac{1}{8}\left(2(\theta+3)(9-\theta)^2\right)^{1/3}, & \mbox{ if } \theta_*\leq \theta < 3\\
1 - \frac{3}{2}\left(\frac{(\theta+1)(\theta-1)^2}{(2\theta-1-\rho)(\theta-2+\rho)(\theta+1+\rho)}\right)^{1/3}, & \mbox{ if } 1<\theta<\theta_*\\
0 & \mbox{ if } \theta=1
\end{array}
\right.
\]
where $\rho=\sqrt{\theta^2-\theta+1}$, and $\theta_*\approx 1.32$ is the 3rd root of the equation
$$3456 - 5184 \theta + 3561 \theta^2 + 596 \theta^3 - 1506 \theta^4 + 100 \theta^5 + \theta^6 =0.$$
\end{theorem}

\begin{figure}[h]\caption{Plot of $R_{\rm max}^{(u)}$ versus $\theta$}\label{fig:rmaxplot}
\begin{center}
\includegraphics[height=2.5in,width=5in,angle=0]{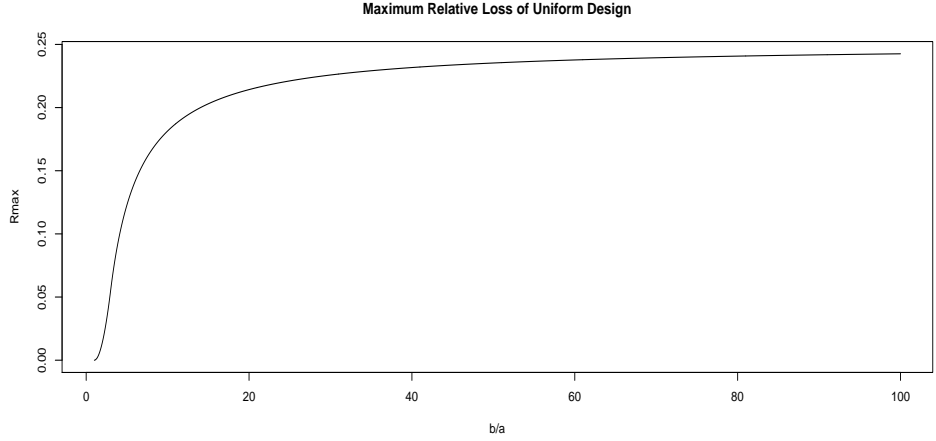}
\end{center}
\end{figure}

\noindent Figure~\ref{fig:rmaxplot} reveals the nature of association between the maximum relative loss of uniform design and the ratio between the upper and lower limits of range of $\mathbf{w}$'s. It can be seen that if the performance of uniform design become worst when $\theta$ is 10 or more, but even in that case $R_{max}^{(u)}$ is less than 1/4. Note that the fact $R_{\rm max}^{(u)}\le 1/4$ was established in Mandal, Yang and Majumdar (2010). Uniform design performs moderately well when $\theta$ lies between 3 and 10, and it performs extremely well if $\theta < 3$.

\subsection{A real example}

\bigskip\noindent The data given in Table~\ref{tab5.1}, reported by Collett (1991), were originally obtained from an experiment conducted at the East Malling Research Station (Hoblyn and Palmer, 1934). The experimenters investigated the vegetative reproduction of plum trees. Cuttings from the roots of older trees of the palms named Common Mussel were taken between October, 1931 and February, 1932. This experiment involved two factors each at two levels. The first factor was time of planting (root stocks were either planted as soon as possible after they were taken, or they were imbedded in sand under cover and were planted in the next spring). The second factor was the length of root cuttings (6 cm or 12 cm). Hoblyn and Palmer used an uniform design and a total of 240 cuttings were taken for each of the four combinations. The response was the condition of each plant (alive or dead) in October, 1932.

\begin{table}[h]\caption{Survival rate of plum root-stock cutting}\label{tab5.1}
\begin{center}
\begin{tabular}{llcc}
\hline
Length of  & Time of   & Number of surviving\\
cutting    & planting  & out of 240\\
\hline
Short      & At once   & 107\\
           & In Spring & 31\\
Long       & At once   & 156\\
           & In Spring & 84\\
\hline
\end{tabular}
\end{center}
\end{table}

\noindent After fitting the logit model, we get, $\hat{\beta}=(-0.5088, -0.5088, 0.7138)^\prime$ and the corresponding ${\mathbf w} = (0.244,0.128,0.221,0.221)^\prime$. If we use this ${\mathbf w}$, the optimal proportions are $\mathbf{p_o}=(0.2818, 0.1686, 0.2748, 0.2748)^\prime$. The corresponding determinant of $X^\prime W X$ is $8.197\times 10^{-3}$. On the other hand, the determinant of the information matrix corresponding to the uniform design is $7.975\times 10^{-3}$.
Thus the uniform design is $(7.975/8.197)^{1/3}=$ 99.1\% efficient. If this was the first of a series of experiments, then the result supports the continued use of the uniform design. Similar calculations with the probit link shows that the uniform design is 99.9\% efficient.

\section{Discussion}

\noindent In this paper and our earlier work, Mandal, Yang and Majumdar (2010), we have investigated various aspects of locally $D$-optimal designs for $2^{2}$ experiments with binary response. Extending these results to experiments with more than two factors is far from a straightforward task. The main difficulty is that a concise expression like $(1)$ is not available. We are currently investigating different methods for the general case. For some special cases, however, the results can be derived with relative ease. For instance, for the $1/2$ fraction of $2^{3}$ given by $C=AB,$ it can be shown that the uniform design is locally $D$-optimal no matter what the assumed values of the parameter may be.

\bigskip
\renewcommand{\theequation}{A.\arabic{equation}}
\appendix

\begin{center}\noindent {\large\bf Appendix}\end{center}

\bigskip \noindent {\textbf{\emph{1. Proof of Theorem~\ref{thm:2}}}}

\bigskip \noindent
To simplify the notations, we use $(v_1, v_2, v_3, v_4)$ instead of $(v_{t1}, v_{t2}, v_{t3}, v_{t4})$ here.
Note that $Q_c(\lambda v_1, \lambda v_2, \lambda v_3, \lambda v_4) = Q_c(v_1, v_2, v_3, v_4)$
for any $\lambda > 0$.
Suppose $0 < a \leq v_i < b$, $i=1,2,3,4$. To minimize $Q_c(v_1, v_2, v_3, v_4)$, we only need to
consider those cases with $v_4 = a$.

\bigskip \noindent If $v_{c4} \geq v_{c1} + v_{c2} + v_{c3}$, then $p_{c4}=0$ and $p_{c1}=p_{c2}=p_{c3}=1/3$.  Thus $q_3 =q_2 =q_1 =0$
and $q_4 =1/27$. Fixing $v_4=a$, it can be verified that
$$
Q_c(v_1, v_2, v_3, a)
=
\frac{a/27}
{v_1 p_2 p_3 p_4 + v_2 p_1 p_3 p_4 + v_3 p_1 p_2 p_4 + a p_1 p_2 p_3}\geq Q_c(b,b,b,a),
$$
where ``$=$" is true if and only if $v_1=v_2=v_3=b$.
In this case,
$$R_{\max}(c)
=1-\frac{1}{3\theta}\left(\frac{9\theta-1}{2}\right)^{2/3},$$
where $\theta=b/a$. Note that $\theta\geq 3$ in this case.

\bigskip \noindent Suppose $v_{c4} < v_{c1} + v_{c2} + v_{c3}$. Then $p_{c1} \geq p_{c2} \geq p_{c3} \geq p_{c4} >0$
and $q_4  \geq q_3  \geq q_2  \geq q_1  > 0$.  Again, we fix $v_4 = a$ and assume
$0 < a \leq v_i \leq b$, $i=1,2,3,4$.

\bigskip \noindent
{\it Case 1:}
If $v_1 \geq v_2 + v_3 + a$, and
then $Q_c(v_1, v_2, v_3, a) \geq  Q_c(b, a, a, a)$,
where ``$=$" is attained if and only if
$v_1 = b$ and $v_2=v_3=v_4=a$.
In this case, $\theta=b/a \geq 3$.

\bigskip \noindent
{\it Case 2:}
If $v_1 < v_2 + v_3 + a$ and $v_1' > v_1$, then $Q_c(v_1', v_2, v_3, a) < Q_c(v_1, v_2, v_3, a)$.

\bigskip \noindent Actually, in this case, $0 < p_1 \leq p_2 \leq p_3 \leq p_4$.
It can be verified that for small enough $\delta > 0$,
$$
Q_c(v_1, v_2, v_3, a)
\geq
\frac{v_1' q_1  + v_2 q_2  + v_3 q_3  + a q_4 }
{v_1' p_2' p_3' p_4' + v_2 p_1' p_3' p_4' + v_3 p_1' p_2' p_4' + a p_1' p_2' p_3'}
\geq
Q_c(v_1', v_2, v_3, a),
$$
where $p_i' = p_i + \delta$, $i=2,3,4$ and $p_1'=p_1 - 3\delta$.

\bigskip \noindent From now on, we only need to consider $Q_c(b, v_2, v_3, a)$ with $b \geq v_2 \geq v_3 \geq a$
and $b < v_2 + v_3 + a$.  In this case, $0 < p_1 \leq p_2 \leq p_3 \leq p_4$.

\bigskip \noindent Similarly, it can be verified that\\
(1) If $b > v_2 > a$, then
$Q_c(b, v_2, a, a) > \min\{Q_c(b, a, a, a),Q_c(b, b, a, a)\}$.\\
(2) If $b > v_3 > a$, then
$Q_c(b, b, v_3, a) > \min\{Q_c(b, b, b, a),Q_c(b, b, a, a)\}$.
\\
(3) If $b> v_2 > v_3 > a$, then
$Q_c(b, v_2, v_3, a) > \min\{Q_c(b, v_2, v_2, a),Q_c(b, v_2, a, a)\}$.\\
(4) If $b> v = v > a$, then
$Q_c(b, v, v, a) > \min\{Q_c(b, b, b, a),Q_c(b, a, a, a)\}$.

\bigskip \noindent In summary, if $v_{c4} < v_{c1} + v_{c2} + v_{c3}$, then
$$Q_c(v_1, v_2, v_3, v_4) \geq \min\{Q_c(b,b,b,a),
Q_c(b,b,a,a), Q_c(b, a, a, a)\}.$$   Based on the proof, the minimum of $Q_c(v_1, v_2, v_3, v_4)$
can only be obtained at $(b, a, a, a)$, $(b, b, a, a)$, or $(b, b, b, a)$.
\hfill{$\Box$}

\vskip 0.5cm
\bigskip \noindent {\textbf{\emph{2. Proof of Theorem~\ref{thm:3}}}}

\bigskip \noindent Given $\theta \geq 1$ and
 $p_{c1}\geq p_{c2} \geq p_{c3} \geq p_{c4}$,
it can be verified that
$$\theta q_1 + q_2 + q_3 + q_4
=\theta p_{c2} p_{c3} p_{c4}
+ p_{c1} p_{c3} p_{c4}
+ p_{c1} p_{c2} p_{c4}
+ p_{c1} p_{c2} p_{c3}
\leq \frac{\theta + 3}{27},
$$
where the second ``$=$" is true if and only if
$p_{c1} = p_{c2} = p_{c3} = p_{c4} = 1/4$.  Similarly,
$\theta q_1 + \theta q_2 +  q_3 + q_4
\leq \frac{2(\theta + 1)}{27},
$
where ``$=$" is true if and only if
$p_{c1} = p_{c2} = p_{c3} = p_{c4} = 1/4$;
$\theta q_1 + \theta q_2 + \theta q_3 + q_4
\leq \frac{3\theta + 1}{27}
$
where ``$=$" is true if and only if
$p_{c1} = p_{c2} = p_{c3} = p_{c4} = 1/4$.
Therefore, if $v_{c4} < v_{c1} + v_{c2} + v_{c3}$,
$$\min\left\{Q_c(b, a, a, a), Q_c(b,b,a,a), Q_c(b,b,b,a)\right\}$$
attains its maximum only at
$p_{c1} = p_{c2} = p_{c3} = p_{c4} = 1/4$.
In other words, the uniform design has smaller $R_{\max}(c)$
than any other design ${\mathbf p}_c$ with $v_{c4} < v_{c1} + v_{c2} + v_{c3}$.

\bigskip \noindent On the other hand, it can be verified that
if $\theta \geq 3$,
the maximum relative loss of uniform
$$R_{\max}^{(u)}
=1-\frac{3}{4}\left(1+\frac{3}{\theta}\right)^{1/3}
> 1-\frac{1}{3\theta}\left(\frac{9\theta-1}{2}\right)^{2/3}
=R_{\max}(c)$$
for any design with $v_{c4} \geq v_{c1} + v_{c2} + v_{c3}$.

\bigskip \noindent In short, $R_{\max}(c)$ attains its
minimum only at the uniform design.
\hfill{$\Box$}

\vskip 0.5cm
\bigskip \noindent {\textbf{\emph{3. Proof of Theorem~\ref{thm:42}}}}

\bigskip\noindent
Let $\theta = \beta/\alpha$ (or $b/a$) $\geq 1$ and $\rho=\sqrt{\theta^2-\theta+1}$. Then
\begin{eqnarray*}
Q_1:=Q(b,a,a,a) &=& Q(\theta,1,1,1) =
\left\{
\begin{array}{ll}
27\left(1+\frac{3}{\theta}\right), & \mbox{ if }\theta\geq 3\\
\frac{(\theta+3)(9-\theta)^2}{4}, & \mbox{ if }1\leq \theta < 3
\end{array}
\right.
\\
Q_2:=Q(b,b,a,a) &=& Q(\theta,\theta,1,1) = \frac{216(\theta+1)(\theta-1)^2}{(2\theta-1-\rho)(\theta-2+\rho)(\theta+1+\rho)}\\
Q_3:=Q(b,b,b,a) &=& Q(\theta,\theta,\theta,1) = \frac{(3\theta+1)(9\theta-1)^2}{4\theta^3}
\end{eqnarray*}
Since $\theta\geq 1$, it can be verified that
$Q_3 \geq Q_2$ and the ``$=$" is true if and only if
$\theta=1$, that is, $a=b$.
Thus $\min\{Q(b,b,b,a), Q(b,b,a,a), Q(b, a, a, a)\} = \min\{Q(b,b,a,a), Q(b, a, a, a)\}$.

\noindent Now we only need to compare $Q_1 = Q(b, a, a, a)$ with
$Q_2 = Q(b, b, a, a)$. It can be verified that
\begin{itemize}
\item[{\bf (i)}] If $\theta=1$, or $\theta=\theta_*$, $Q_1=Q_2$.
\item[{\bf (ii)}] If $1 < \theta < \theta_*$, $Q_1 > Q_2$.
\item[{\bf (iii)}] If $\theta > \theta_*$, then $Q_1 < Q_2$.
\end{itemize}
Here
$\theta_*\approx 1.32$ is the 3rd root of equation
$$3456 - 5184 \theta + 3561 \theta^2 + 596 \theta^3 - 1506 \theta^4 + 100 \theta^5 + \theta^6 =0~.$$
Then $R_{\rm max}^{(u)}$ can be obtained accordingly.
\hfill{$\Box$}

\end{document}